\documentclass[12pt]{article}
\begin{document} 
\title{ Relativistic Elasticity}

\author{Robert Beig\\Institut f\"ur Theoretische Physik der Universit\"at
Wien\\ Boltzmanngasse 5, A-1090 Vienna, Austria\\[1cm] Bernd G. Schmidt\\
Max-Planck-Institut f\"ur Gravitationsphysik\\ Albert-Einstein-Institut\\
Am M\"uhlenberg 1, D-14476 Golm, Germany}
\maketitle

\begin{abstract}
{Relativistic elasticity on an arbitrary
spacetime is formulated as a Lagrangian field theory which is covariant
under spacetime diffeomorphisms. This theory is the relativistic 
version of classical elasticity  in the hyperelastic, materially
frame-indifferent  case and, on Minkowski space, reduces to the latter
in the limit $c\to\infty$. The  field equations are cast 
into a first -- order
symmetric hyperbolic system. As a consequence one obtains local--in--time 
existence and uniqueness theorems under various circumstances.}


\end{abstract}

\renewcommand{\theequation}{\thesection.\arabic{equation}}

\section{{\bf Introduction}}

By far the most popular relativistic matter model has been that of a
perfect fluid. One reason for this lies in the relative simplicity
of this model. Another reason is that astrophysical objects,
which are the objects most likely to exhibit effects of Special or
General Relativity, are usually described quite well as bodies
composed of perfect fluid. Elasticity, on the other hand, is
relevant for describing the statics and dynamics of solids which we
meet in everyday life, and for this relativistic effects can safely be
ignored. But, whatever experimental relevance it might have (e.g. for
describing neutron star crusts or bar detectors of gravitational waves),
it is clearly of conceptual interest to have a consistent theory of bulk
matter which is in accordance with the principles of relativistic physics.

A basic consistency check on such a theory is to see whether its
equations have a well--posed initial value problem.
Christodoulou, in the book \cite{Ch}, has announced without proof a
general existence theorem, covering the case of elasticity, using a
new notion of hyperbolicity. We show that the elasticity equations on
an arbitrary spacetime can in a very natural manner be cast into the
form of a quasilinear, symmetric hyperbolic system\footnote{For 
nonrelativistic elasticity such a formulation has been found 
by John \cite{Jo}.}.

Relativistic elasticity has been treated by many authors. The
earliest references we are aware of are Herglotz \cite{He}, in 1911(!) for
Special Relativity and Nordstr\"om \cite{No} in 1916, on General Relativity.
A very influential paper has been that of Carter and Quintana \cite{CaQ}, of
which a clear presentation has been given by Ehlers \cite{Eh}. A very
recent application of the Carter formalism is by Karlovini and Samuelsson
\cite{Karl}. We also would like to mention the work of Maugin
\cite{Mau}. An excellent exposition can be found in the text book of Soper
\cite{So}.      Recent useful references are  Kijowski and Magli
\cite{KiM}, which also contains an extensive bibliography, Christodoulou
\cite{Ch} and Tahvildar--Zadeh \cite{Ta}.

In the present work we treat elasticity as a field theory derived from a
Lagrangian. Our basic fields are maps from spacetime into a 3--dimensional
"material manifold'' $\mathcal B$. Objects defined on $\mathcal B$ are 
physically interpreted
as properties of the material prior to the action of deformations or other
fields.

In Section 2 we review general properties of Lagrangian field
theories with Lagrangians depending on the map $f$ and its first
derivatives and which are covariant under spacetime diffeomorphisms. 
In Section 3 we show that 
the Euler--Lagrange equations of these theories, when written as a
first-order system, can in a very natural manner be rewritten as a
"symmetric system", i.e. as a quasilinear first -- order system where the
coefficient matrices of the derivative -- terms are symmetric.

Now recall that, in the quasilinear case, the coefficient matrices are in
general functions of the independent and dependent variables. The former
are points in spacetime. The latter variables, in our case, consist of 
possible
values for the field and its first derivatives. All these together form
what we call the "deformation bundle" $\mathcal D$ over spacetime.
According to the general theory the symmetric system found in Sect.3  is
symmetric hyperbolic at a point of  $\mathcal D$ if there is a covector at
the corresponding spacetime  point so that a certain positivity
requirement is fulfilled for the coefficient matrices. The existence
of such ``hyperbolic points'' of $\mathcal D$ requires further
constitutive hypotheses to which we turn in Sect.4.
Here we define a class of theories for which $\mathcal D$ has a
subbundle consisting of ``unstrained'' or ``natural'' points
which are in addition stress -- free and for which the elasticity 
tensor has the standard "Hookean"
structure. We show that, for such materials, our equations are indeed
symmetric hyperbolic near natural points in the deformation bundle. 

In Section 5 we discuss various physical sitations covered by our
existence theorem and the interpretation of the initial data.

Finally (in Appendix B) we treat the non--relativistic limit $c\to\infty$
of elasticity in Minkowski space. As to be expected, the basic
structure of the equations remains essentially unchanged and we
obtain again local existence and uniqueness. These equations are
equivalent to those of the standard nonrelativistic theory 
in the hyperelastic, materially frame-indifferent case (see
Ciarlet \cite{Ci} and Gurtin \cite{Gu}).

\section{{\bf Preliminaries}}
\setcounter{equation}{0}

States of relativistic continua are best described by maps $f$ from
spacetime $(M,g)$ to a 3--dimensional manifold $\mathcal B$, called
"material manifold". In local coordinates we write $X^A=f^A(x^\mu)$. The
space $\mathcal B$ is the abstract collection of particles making up the
continuous medium. By "abstract" we mean that we view $\mathcal B$ as 
a space of
its own right independent of "physical space". There are several reasons
for doing so, the most important one being that there is, in a
relativistic theory, no natural notion of "spatial location at given
time".

The inverse images of points of $\mathcal B$ are supposed to form a timelike
congruence of $M$. Thus there is, given $f$,  a timelike vector $u^\mu$, 
unique up to scale, so that
\begin{equation}
  \label{1} 
u^{\mu} \partial_{\mu}f^A=0\ ,
\end{equation}
Given a time orientation for $(M,g)$, $u^\mu$ is fixed by requiring
$u^\mu$ to be future--pointing and 
\begin{equation}
  \label{2} 
 g_{\mu\nu}u^\mu u^\nu=-1\ .
\end{equation}
The dynamics will be given by an action principle of the forme
\begin{equation}
  \label{3}
S[f]=\int_M\rho(f,\partial f;g)\sqrt{-det(g)}\ d^4x  
\end{equation}
with associated Euler -- Lagrange equations 
\begin{equation}
  \label{4}
-\mathcal E_A:=\frac{1}{\sqrt{-det(g)}}\partial_{\mu}\left(\sqrt{-det(g)}
\frac{\partial\rho}{\partial(\partial_\mu f^A)}\right)-\frac{\partial\rho}
{\partial f^A}=0\ .
\end{equation}
The Lagrangian $\rho$ is thus a function on the "deformation bundle"
$\mathcal D$ with base $M \times \mathcal B$, 
coordinatized by $(x^\mu,X^A,F^B{}_\nu)$  
where $F^A{}_\mu$ is of rank three
with null space timelike w.r. to $g$. The geometrical interpretation of
$F^A{}_\mu$ at $(x^\mu, X^A)$ is that of a linear map from $T_x(M)$ to
$T_X(\mathcal B)$.     While we have no desire to enter formal excesses, 
we do from
now on use notation which emphasizes the difference between a general fibre
point 
$(X^A,F^B{}_\mu)$ and a pair of values $(f^A(x),\partial_\mu f^B{}(x))$ of
$(X^A,F^B{}_\nu)$ along a particular map $f$. The four velocity $u^\mu$ for
example can be viewed as a function on the deformation bundle. A convenient
way of expressing the rank--condition and explicitly writing down this
function
is as follows: Choose a volume form $\Omega_{ABC}$ on $B$ and require 
\begin{equation}
  \label{5}
\omega_{\mu\nu\lambda}=F^A{}_\mu F^B{}_\nu F^C{}_\lambda \Omega_{ABC}(X)
\end{equation}
to be non--zero and 
\begin{equation}
  \label{6}
\epsilon^{\mu\nu\lambda\rho}\omega_{\nu\lambda\rho}
\end{equation}
to be future--pointing timelike, where $\epsilon^{\mu\nu\lambda\rho}$ is
the volume form associated with the spacetime metric. This means that the
linear maps
$F^A{}_\mu$ preserve orientation. Given this orientation, there is a
positive function
$n$ on the deformation bundle so that 
\begin{equation}
  \label{7}
u^\mu={1\over 3!n}\epsilon^{\mu\nu\lambda\rho}\omega_{\nu\lambda\rho}
\end{equation}
is normalized. 

Along a map $f$, $nu^\mu$ is a vector field on $M$ satifying the
continuity equation
\begin{equation}
  \label{8}
\nabla_\mu(nu^\mu)=0
\end{equation}
identically. The physical meaning of $n$ is that of a density of number
of particles in the state given by $f$. 

We now come back to the Lagrangian $\rho$ which is a function on the
deformation bundle with $x$-dependence only via the spacetime metric $g$,
i.e. $\rho=\rho(X,F;g)$.  As a further ingredient we assume covariance of
the Lagrangian under spacetime diffeomorphisms, i.e. that, along a map $f$,
$\rho$ behaves as a scalar under transformations of the coordinates
$x^\mu$. This is equivalent to the following condition on $\rho$:
\begin{equation}
  \label{9}
\mathcal{L}_\xi\rho={\partial\rho\over\partial X^A} 
(\mathcal L_\xi f)^A
+{\partial\rho\over\partial F^A{}_{\mu}}
\mathcal L_\xi (\partial_\mu f^A)
+{\partial\rho\over\partial g^{\mu\nu}}(\mathcal L_\xi g)^{\mu\nu}
\end{equation}
for arbitrary vector fields $\xi^\mu(x)$. Diffeomorphism invariance has
several important consequences.
The first is that the energy momentum tensor $T_{\mu\nu}$ defined by 
\begin{equation}
  \label{10}
T_{\mu\nu}=2{\partial\rho\over\partial
 g^{\mu\nu}}-\rho g_{\mu\nu}
\end{equation}
obeys the identity
\begin{equation}
  \label{11}
-\nabla_\nu T_\mu{}^\nu=\mathcal E_AF^A{}_\mu\ ,
\end{equation}
independently of the field equations. (Notice that both the left--hand side
of (\ref{11}) and $\mathcal E_A$ should be viewed as functions on the 
"second jet
space" with elements given by $(x^\mu,X^A,F^A{}_\nu,
F^B{}_{\lambda\rho}=F^B{}_{(\lambda\rho)})$). From Equ. (\ref{11})
and the regularity condition on $F^A_\mu$ it follows
that the Euler-Lagrange equations are equivalent to the
energy-momentum tensor being divergence-free.

Secondly, writing out the Lie--derivatives in (\ref{9}) in terms of
partial derivatives, we find the equivalent
relation
\begin{equation}
  \label{12}
F^A{}_\mu{\partial\rho\over
\partial F^A{}_\nu}=2g^{\nu\lambda}{\partial\rho\over\partial 
g^{\mu\lambda}}\ .
\end{equation}
Equ.(\ref{12}), in turn, has the corollaries
\begin{equation}
  \label{13}
{\partial\rho\over\partial g^{\mu\nu}}u^\nu=0
\end{equation}
and
\begin{equation}
  \label{14}
F^A{}_{[\mu} g_{\nu]\lambda}{\partial \rho\over\partial
F^A{}_\lambda}=0\ ,
\end{equation}
Equ.(\ref{13}) implies that
\begin{equation}
  \label{15}
T_{\mu\nu}u^\nu=-\rho u_\mu\ .
\end{equation}
Thus the energy momentum tensor has the form
\begin{equation}
  \label{16'}
T_{\mu\nu}=\rho u_\mu u_\nu +S_{\mu\nu}\ ,
\end{equation}
where the stress tensor satisfies $S_{\mu\nu}u^\nu=0$. It follows
that there exists
$\tau_{AB}=\tau_{(AB)}$ on the  deformation bundle so that
\begin{equation}
  \label{16}
T_{\mu\nu}=\rho u_\mu u_\nu +n\tau_{AB} F^A
{}_{\mu}F^B{}_{\nu}\ .
\end{equation}
(The factor $n$ in front of $\tau_{AB}$ is inserted for later convenience.)
In other words, in the rest frame of matter, $T_{\mu\nu}$ has no mixed
space--time components and the Lagrangian $\rho$ is the energy density of
the matter. The quantity $\tau_{AB}$ 
is called "second Piola--Kirchhoff" stress tensor in the elasticity
literature.

As a special case consider a Lagrangian which just depends on $n$, as
defined above. We find that 
\begin{equation}
  \label{17}
{\partial n\over\partial g^{\mu\nu}}={n\over2} h_{\mu\nu}
\end{equation}
with
\begin{equation}
  \label{18}
h_{\mu\nu}=g_{\mu\nu}+u_\mu u_\nu\ .
\end{equation}
Thus
\begin{equation}
  \label{19}
T_{\mu\nu}=\rho u_\mu u_\nu+p h_{\mu\nu}
\end{equation}
with 
\begin{equation}
  \label{20}
p=n{\partial\rho\over\partial n}-\rho\ .
\end{equation}
Since, by (\ref{11}), the field equations on $f$ are equivalent to
$\nabla_\nu T_\mu{}^\nu=0$, we have recovered the standard formulation of
perfect fluids: one simply considers $\nabla_\nu T_\mu{}^\nu=0$ as the
equations of motion for $u^\mu$ as a vector field on $M$, supplements them
by the continuity equation (\ref{8}) and completely
forgets about $f:M\to \mathcal B$.

\section{{\bf The symmetric system}}
\setcounter{equation}{0}

The equations $\mathcal E_A=0$ can be written as (we suppress the
$g_{\mu\nu}$--dependence)
\begin{equation}
  \label{3.1}
M^{\mu\nu}{}_{AB}(f,\partial f)\partial_\mu\partial_\nu
f^B(x)=G_A(f,\partial f)\ ,
\end{equation}
where 
\begin{equation}
  \label{3.2}
M^{\mu\nu}{}_{AB}:= {\partial^2\rho\over\partial F^B{}_{\nu}\partial
F^A{}_{\mu}}= M^{\nu\mu}{}_{BA} 
\end{equation}
The quantities $M^{\mu\nu}{}_{AB}$ can be viewed as a quadratic form on
the space of matrices $m^A{}_\mu$ at each point of the deformation bundle.
Alternatively they can be viewed as a map sending matrices $m^A{}_\mu$ to
matrices $n_B{}^\nu$. This map, the "Legendre map", is in what follows
required to be non--degenerate, i.e. $M^{\mu\nu}{}_{AB} m^B{}_\nu=0$
implies $m^A{}_\mu=0$. Another restriction on $M^{\mu\nu}{}_{AB}$
follows from Equ.(\ref{14}). Namely, differentiating  (\ref{14}) w.r. to
$F^A{}_\mu$ and using the rank--condition  one easily finds
that
\begin{equation}
  \label{3.3}
u_\mu u^{[\nu}M^{\lambda]\mu}{}_{AB}=0\ .
\end{equation}
We now have the properties of $M^{\mu\nu}{}_{AB}$ necessary for
rewriting (\ref{3.1}) as an equivalent first--order system. Let us define
\begin{equation}
  \label{3.4}
W^{\mu\nu}{}_{AB}{}^{(\lambda)}:= u^\mu
M^{\lambda\nu}{}_{AB}-2u^{[\lambda}M^{\nu]\mu}{}_{BA}
$$
$$
=u^\mu
M^{\lambda\nu}{}_{AB}+u^\nu M^{\lambda\mu}{}_{BA}-u^\lambda
M^{\nu\mu}{}_{BA}\ .
\end{equation}
By (\ref{3.2})
\begin{equation}
  \label{3.6}
W^{\mu\nu}{}_{AB}{}^{(\lambda)}=W^{\nu\mu}{}_{BA}{}^{(\lambda)}
\end{equation}
Furthermore, since the last two terms in (\ref{3.4}) are antisymmetric
in
$\lambda$ and $\nu$, from (\ref{3.1}) there follows that
\begin{equation}
  \label{3.5}
W^{\mu\nu}{}_{AB}{}^{(\lambda)}(f,\partial f)\partial_\lambda\partial_\nu
f^B=u^\mu(f,\partial f) G_A(f,\partial f)
\end{equation}
We now replace (\ref{3.1}) by the following first--order system:
\begin{equation}
  \label{3.7}
W^{\mu\nu}{}_{AB}{}^{(\lambda)}(X,F)\partial_\lambda
F^B{}_\nu=u^\mu(X,F) G_A(X,F)
\end{equation}
\begin{equation}
  \label{3.8}
-u^\lambda(X,F)\partial_\lambda X^A=0\ .
\end{equation}
We can write (\ref{3.7},\ref{3.8}) in matrix form
as 
\begin{equation}
  \label{S}
\left(\matrix{W^{\mu\nu}{}_{AB}{}^{(\lambda)}&0\cr
0&-\delta^A{}_B\ u^\lambda\cr}\right) 
\partial_\lambda\left(\matrix{F^B{}_\nu\cr
X^B}\right)=\left(\matrix{.\cr .}\right)
\end{equation} 
and see that we have a symmetric system because of (\ref{3.6})
\footnote{The quantity $W^{\mu\nu}{}_{AB}{}^{(\lambda)}$ also appears in
\cite{Ch}, but is used there for a different purpose.}.

We have just seen that, taking a solutions
$X^A = f^A(x)$ of (\ref{3.1})  and setting 
$F^A{}_\mu=\partial_\mu f^A(x)$ we obtain a solution of
(\ref{3.7},\ref{3.8}). The converse is given by the following 

{\bf Theorem 1:}
Suppose we have a solution pair  ($X^A(x), F^A{}_\mu(x)$) of Equ.'s
(\ref{3.7},\ref{3.8}) and a hypersurface $\Sigma$, on which 

(i) $\partial_iX^A{}$ has  rank three 
\footnote{Condition (i) is the same as saying that $u^\mu(X,F;g)$ is
transversal to $\Sigma$}

(ii)
 $
(\partial_i X^A - F^A{}_i)|_\Sigma=0$

Then $f^A = X^A(x)$ satisfies $\partial_\mu f^A=F^A{}_\mu$ and Equ.'s
(\ref{3.1}).

{\bf Proof:}
 Contraction of (\ref{3.7}) with $u_\mu$ and 
 (\ref{3.3}) imply
\begin{equation}
  \label{3.9}
M^{\lambda\nu}{}_{AB}\partial_\lambda F^B{}_\nu=G_A
\end{equation}
It remains to show that $F^A{}_\mu=\partial_\mu f^A$. 
Inserting (\ref{3.9}) back into  (\ref{3.5}) gives 
\begin{equation}
  \label{3.10}
u^\lambda M^{\nu\mu}{}_{BA}\partial_{[\lambda}F^B{}_{\nu]}=0\ .
\end{equation}
Using the regularity of the Legendre map this leads to 
\begin{equation}
  \label{3.11}
u^\lambda \partial_{[\lambda}F^B{}_{\nu]}=0\ .
\end{equation}
We now  write
$X=(X^A)$ as a triple of 0--forms and $F=F^A{}_\mu dx^\mu$ as a
triple of 1--forms. 
Consider the expression $F-dX$. From the definition of $u^\mu$ and
(\ref{3.8})  it follows that
\begin{equation}
  \label{3.12}
i_u (F-dX)=0
\end{equation}
Thus
\begin{equation}
  \label{25}
\mathcal L_u(F-dX)=i_u d(F-dX)=i_u dF=0\ ,
\end{equation}
where we use (\ref{3.11}) in the last equality.  The condition (ii)
states that the pull--back of $F-dX$ to $\Sigma$ is zero. Thus, since
$u^\mu$ is transversal to $\Sigma$, $F-dX$ vanishes identically,
and the proof
is complete.

Before commenting on the relation between the first and second--order
systems we introduce a further piece of notation. The rank-condition on
$F^A{}_\mu$ means that it has an inverse on the space orthogonal to
$u^\mu$. Thus there exists $F_A{}^\mu$  such that 
\begin{equation}
  \label{3.13}
F_A{}^\mu F^B{}_\mu=\delta_A^B\ ,\ \ \  F^A{}_\mu F_A{}^\nu = h_\mu^\nu\ ,
 \end{equation}
where $h_\mu^\nu$ is defined by (\ref{18}).

Defining 
\begin{equation}
  \label{3.15}
H^{AB}=F^A{}_\mu F^B{}_\nu g^{\mu\nu}\ .
\end{equation}
we can write $F_A^\mu$ down explicitly as
\begin{equation}
  \label{3.14}
F_A{}^\mu=H_{AB}F^{B\mu}=H_{AB}g^{\mu\nu} F^B{}_\nu\ ,
\end{equation}
where $H_{AB}$ is defined as the inverse of $H^{AB}$, i.e.
$H_{AB}H^{BC} = \delta^C_A$\footnote{$H_{AB}$ describes the
distance of particles of $\mathcal B$ in spacetime.}. 
  
With these definitions the relation (\ref{3.3}) implies the existence of
quantities
$\mu_{AB}=\mu_{(AB)}$ and $U_{CADB}=U_{DBCA}$ such that
\begin{equation}
  \label{3.16}
M^{\mu\nu}{}_{AB}=-\mu_{AB}u^\mu u^\nu +U_{ACBD}F^{C\mu}F^{D\nu}
\end{equation}

Thus the regularity conditions is equivalent to both $\mu_{AB}$ and
$U_{CADB}$ being regular in that $\mu_{AB}\alpha^B=0\Longrightarrow
\alpha^A=0$ and $U_{ABCD}\alpha^{CD}=0\Longrightarrow
\alpha^{AB}=0$.

Next, using (\ref{3.3}) and (\ref{3.4}), it follows that 
\begin{equation}
  \label{3.18}
W^{\mu\nu}{}_{AB}{}^{(\lambda)}u_\lambda=\mu_{AB} u^\mu u^\nu +
U_{ACBD}F^{C\mu}F^{D\nu}\ .
\end{equation}
Note that, compared to $M^{\mu\nu}{}_{AB}$, only the sign of the first
term has changed. The regularity of $M^{\mu\nu}{}_{AB}$ is thus
equivalent to the requirement that $u_\mu$ be a non--characteristic
covector for the first--order system (\ref{3.7},\ref{3.8}).

Now recall the notion of characteristic covectors for the second--order
system (\ref{3.1}). Namely $k_\mu$ is characteristic iff 
\begin{equation}
  \label{3.19}
\Delta_{AB}=M^{\mu\nu}{}_{AB}k_\mu k_\nu
\end{equation}
is singular, i.e. $\Delta=det(\Delta_{AB})=0$. If $k_\mu$ is
characteristic in this sense and $\alpha^A$ an associated eigenvector,
i.e. $\Delta_{AB}\alpha^B=0$, then
$W^{\mu\nu}{}_{AB}{}^{(\lambda)}k_\lambda k_\nu\alpha^B=0$ i.e. $k_\mu$
is also characteristic for the system (\ref{3.7},\ref{3.8}).

The first--order system (\ref{3.7},\ref{3.8})  is called symmetric
hyperbolic if, in addition to the symmetry of the coefficients, there
exists a "subcharacteristic" covector (hypersurface element), 
that--is--to--say a covector $k_\lambda$
 for which the matrix
\begin{equation}
  \label{matrix}
 \left(\matrix{W^{\mu\nu}{}_{AB}{}^{(\lambda)}k_\lambda&0\cr
0&-H_{AB}\ u^\lambda k_\lambda\cr}\right) 
\end{equation}
is positive definite in the variables $(m^A{}_\mu,l^A)$. In fact, using the
variables
$m^A{}_\mu u^\mu=-\alpha^A$, $m^A{}_\mu F^{B\mu}=\alpha^{AB}$, $l^A$, the
quadratic form corresponding to (\ref{matrix}), for $u_\lambda=k_\lambda$, is
given by
\begin{equation}
  \label{3.20}
\mu_{AB}\alpha^A\alpha^B
+U_{ACBD}\alpha^{AC}\alpha^{BD}+H_{AB}l^Al^B\ .
\end{equation}
Therefore, $u_\lambda$ is subcharacteristic for (\ref{3.7},\ref{3.8}) iff the
first two forms in (\ref{3.20}) are positive definite.  Notice that these
conditions are also sufficient for regularity. The validity of these
conditions is studied in the next section.

\section{Hyperbolicity}
\setcounter{equation}{0}
We first return to the covariance condition Equ.(\ref{12}) and claim that 
it is equivalent to the requirement that $\rho(X,F;g)$ is a function
just of $(X,H)$ with $H^{AB}$ given by (\ref{3.15}). To prove this assertion
one first observes that Equ.(\ref{13}) implies
\begin{equation}
 \label{4.111}
\rho(X^A, F^B{}_\mu; g_{\nu \lambda}) = 
\rho(X^A, F^B{}_\mu; g_{\nu \lambda} + s u_\nu u_\lambda)
\end{equation}
for all real numbers $s$. Setting $s=1$ in Equ.(\ref{4.111})
 and using that
$h_{\mu \nu} = F^A{}_\mu F^B{}_\nu H_{AB}$ we infer that
there is a function $\sigma$ so that
\begin{equation}
 \label{4.112}
\rho(X^A,F^B{}_\mu;g_{\nu \lambda}) = 
\sigma(X^A,F^B{}_\mu,H^{CD})\ .
\end{equation}
Using (\ref{4.112}) again in Equ.(\ref{12}) we see that $\sigma$ has to be
independent of $F^A{}_\mu$, thus proving our assertion.

We now write $\rho$ as
\begin{equation}
  \label{4.1}
\rho=n\epsilon\ ,
\end{equation}
where $\epsilon$ is a positive function with
\begin{equation}
  \label{4.2}
\epsilon=\epsilon(X^A,H^{BC})\ .
\end{equation}
The quantity $\epsilon$ is the relativistic version of the
``stored-energy function'' of standard elasticity. It is possible
to factorize $\rho$ in this way
since the number density $n$ is uniquely given in terms of 
$(X^A,H^{BC})$ by virtue of
\begin{equation}
  \label{4.3}
6n^2=H^{AA'}H^{BB'}H^{CC'}\Omega_{ABC}\Omega_{A'B'C'}\ ,\ n>0\ .
\end{equation}
Just as a pair $(X^A,F^B{}_\nu)$, along a map $f:M\to \mathcal B$, measures
deformation, the pairs $(X^A, H^{BC})$ measure "strain". Thus triples
$(x,X,H)$ might be considered as forming the "strain bundle" over
$M \times \mathcal B$ in which the deformation bundle $\mathcal D$ 
is embedded via Equ.(\ref{3.15}).
Using the definition of $F_A{}^\mu$ we easily find that 
\begin{equation}
  \label{4.4}
{\partial F_A{}^\mu\over\partial F^B{}_\nu}=-F_A{}^\nu
F_B{}^\mu-H_{AB}u^\mu u^\nu
\end{equation}
and, from the definition of $u^\mu$, that
\begin{equation}
  \label{4.5}
{\partial u^\mu\over\partial F^A{}_\nu}=-F_A{}^\mu u^\nu\ . 
\end{equation}
From (\ref{4.3}) we deduce that 
\begin{equation}
  \label{4.6}
{\partial n\over\partial F^A{}_\mu}=n F_A{}^\mu
\end{equation}
which, using (\ref{4.4},\ref{4.5}), leads to
\begin{equation}
  \label{4.7}
{\partial^2n\over\partial F^B{}_\nu\partial F^A{}_\mu}=2n F_A{}^{[\mu}
F_B{}^{\nu ]}-nH_{AB}u^\mu u^\nu\ .
\end{equation}
The quantities $\mu_{AB}$ and $U_{ACBD}$ entering the expression
(\ref{3.16}) for $M^{\mu\nu}{}_{AB}$ using (\ref{4.6}, \ref{4.7}) can now
be written as 
\begin{equation}
  \label{4.8}
\mu_{AB}=n(\epsilon H_{AB}+\tau_{AB})
\end{equation}
\begin{equation}
  \label{4.9}
U_{ACBD}=n(\tau_{AB}H_{CD}+\tau_{AC}H_{BD}+\tau_{BD}H_{AC}
+2{\partial\tau_{BD}\over\partial H^{AC}}+2\epsilon H_{A[C}H_{D]B})\ ,
\end{equation}
where the second Piola--Kirchhoff stress tensor $\tau_{AB}$
entering (\ref{16}) is given by
\begin{equation}
  \label{4.10}
\tau_{AB}=2{\partial\epsilon\over\partial H^{AB}}\ .
\end{equation}
Note that $\mu_{AB}$ and $U_{ACBD}$ are functions solely on the strain
bundle. Furthermore the dependence on the volume form
$\Omega_{ABC}$  is only via the multiplicative factor $n$. It will
follow that our results in this paper do not depend on the choice of
$\Omega_{ABC}$.

Of particular interest are points
$(\stackrel{\circ}{x},\stackrel{\circ}{X},\stackrel{\circ}H)$ of the
strain bundle which are stress--free i.e. for which $\tau_{AB}$
vanishes. Then the regularity condition on $\mu_{AB}$ is clearly
satisfied since $\epsilon>0$, but that on $U_{ACBD}$ becomes delicate:
Since the fourth term in (\ref{4.9}), in addition to $U_{ACBD}=U_{CADB}$,
is symmetric in both $(BD)$ and $(AC)$, this term annihilates elements
$\alpha^{AB} $ with $\alpha^{AB}=\alpha^{(AB)}$  which the last term
simply multiplies by a positive constant. On the other hand the second
term annihilates elements $\alpha^{AB}$ with $\alpha^{AB}=\alpha^{[AB]}$.
Thus regularity requires a balance between these two terms, which is not
necessarily satisfied in the cases one wants to consider. But there is a
possible cure: One can try to add further terms proportional to
$H_{C[A}H_{B]D}$ to $U_{CABD}$ so that regularity is satisfied. By that
same token one might hope to also satisfy part 2 of the hyperbolicity 
requirement, namely the existence of a  timelike covector $k_\mu$. We
would, by this manouvre, change $M^{\mu\nu}{}_{AB}$ into
$\bar M^{\mu\nu}{}_{AB}$ and accordingly
$W^{\mu\nu}{}_{AB}{}^{(\lambda)}$ into $\bar
W^{\mu\nu}{}_{AB}{}^{(\lambda)}$. The second-order system, 
however, would be unchanged. We would thus be able to use $\bar
W^{\mu \nu}{}_{AB}{}^{(\lambda)}$ instead of
$W^{\mu \nu}{}_{AB}{}^{(\lambda)}$ in Theorem 1. Consequently
the statements at the end of Section 3 can be
generalized by saying: The system (\ref{3.7},\ref{3.8}) with   $\bar
W^{\mu\nu}{}_{AB}{}^{(\lambda)}$ is symmetric hyperbolic if there exist
quantities $\Lambda_{ACBD}$ on the strain bundle with
$\Lambda_{ACBD}=\Lambda_{CADB}$, but also $\Lambda_{ACBD}=-\Lambda_{ADBC}$,
in such a way that 
\begin{equation}
  \label{4.11}
\bar U_{ACBD}=U_{ACBD}+\Lambda_{ACBD}
\end{equation}
 is positive definite. A necessary condition for this to be the case is
that $U_{ACBD}$ is positive definite on rank--one elements, i.e. on elements
$\alpha^{AC}$ of the form
\begin{equation}
  \label{4.12}
\alpha^{AC}=\lambda^A\mu^C\ ,
\end{equation}
since that property remains unchanged under (\ref{4.11}). Thus the question
is whether rank--one positivity, called Legendre--Hadamard condition in the
time independent theory,  is sufficient for the existence of a suitable
$\Lambda_{ABCD}$. The answer, in three or more space dimensions, is in
general "no" (see Ball \cite{Ba} and references therein). 
Luckily, in the situation we
shall presently consider, the answer is affirmative 
\footnote{Christodoulou claims that his new hyperbolic theory
just requires rank-one positivity (see \cite{Ch}). The standard
theorems for nonrelativistic elasticity in the second-order formulation
also only require rank-one positivity, see \cite{Hu}.}.

We firstly imagine there to be given a positive definite metric
$G^{AB}(X)$ (usually taken flat in standard elasticity) 
on $\mathcal B$ playing the role of ``zero strain''.
Points $(\stackrel{\circ}{x},\stackrel{\circ}{X},\stackrel{\circ}{F})$ 
on $\mathcal D$ will be called ``natural'' or strain-free when 
$\stackrel{\circ}{F}\!^A{}_\mu \stackrel{\circ}{F}\!^B{}_\nu 
g^{\mu\nu}(\stackrel{\circ}{x})=G^{AB}(\stackrel{\circ}{X})$
\footnote{We note that there is  for general spacetimes no cross section
$X^A = f^A(x), F^A{}_\mu(x) = (\partial_\mu f^A)(x)$ consisting 
of natural points. The existence of such a map $f: M \rightarrow \mathcal B$
requires the flow of the associated vector field $u^\mu$ on $M$ 
to be Born-rigid (see \cite{Pi}).}.  
Next suppose that the stored energy function $\epsilon$
depends on $X$ only via $G^{AB}$ and is covariant under diffeomorphisms
of $\mathcal B$. It is not difficult to see that this
implies that $\epsilon$ only depends on the principal invariants
of the linear map $\mathcal H$ with components given by 
$\mathcal{(H)}^A{}_B = H^{AC}G_{CB}(X)$. 
{In that case it is easy
to see that $M^{\mu\nu}{}_{AB}$ transforms tensorially both under
transformations of spacetime and body coordinates, and in that case the
full field equations can be written
$M^{\mu\nu}{}_{AB}\bar\nabla_\mu\nabla_\nu f^B=0$, where
$\bar\nabla_\mu$ is the appropriate "double--covariant" derivative.}
(We could, without altering our results, add some thermodynamics by
allowing $\epsilon$ to depend on an additional scalar function on 
$\mathcal B$, the entropy density.) 

Finally we assume
that the stress $\tau_{AB}$ vanishes at natural points. We now
expand $\epsilon$ at natural points or, equivalently, 
at $\mathcal H = \mathcal I$, where $\mathcal I$ is the identity map.
It follows that there are constants $m,q,r$  
such that
\begin{equation}
  \label{4.13}
\epsilon=m+{m\over
8}[q\  tr(\mathcal H - \mathcal I)^2 + 
r (tr(\mathcal H - \mathcal I))^2]
+ O((\mathcal H - \mathcal I)^3)
\end{equation}
where $m>0$ is the rest mass per particle. (In the prestressed case there
would, in Equ. (\ref{4.13}), appear a linear term of the form  $p
(2\!\stackrel{\circ}n)^{-1}(tr(\mathcal H -
\mathcal I))$. If the background pressure $p$  is positive our results
below continue to hold but the expressions for the phase velocities
change.)

The connection between
$q,r$ and the Lam\'e ``constants'' is
$\lambda=m\stackrel{\circ}{n}q$ and $\mu=m\stackrel{\circ}{n}r$. \footnote{Since
in our formulation the volume element $\Omega$ is 
independent of the metric $G^{AB}$, there is in general no need 
for  $\stackrel{\circ}{n}$ to be constant. Thus the Lam\'e coefficients
could depend on $\stackrel{\circ}{X}$, but not their quotient. The
phase velocities of sound are constant, as we see below.} 

From (\ref{4.13}) and  (\ref{4.9}) we deduce
\begin{equation}
  \label{4.14}
\stackrel{\circ}{U}_{ACBD}=m\stackrel{\circ}{n}\left[
qG_{AC}G_{BD}
+2rG_{D(C} G_{A)B} 
+2 G_{A[C}G_{D]B}
\right] 
\end{equation}
One easily finds that $\stackrel{\circ}{U}$ is rank--one positive iff
$r>0$ and $2r+q>0$. We set 
\begin{equation}
  \label{4.15}
q=c_1^2-2c_2^2\ ,\ \ r=c_2^2\ .
\end{equation}
Inserting (\ref{4.14}) and $\stackrel{\circ}{\tau}_{AB}=0$ into
$M^{\mu\nu}{}_{AB}$ we find that
$\Delta=det(\Delta_{AB})=M^{\mu\nu}{}_{AB}k_\mu k_\nu$, in a frame
adapted to $\stackrel{\circ}{u}{}\!^\mu$, is given by 
\begin{equation}
  \label{4.16}
(m \stackrel{\circ}{n})^3(\stackrel{1}{g}{}\!^{\mu\nu}k_\mu k_\nu)
(\stackrel{2}{g}{}\!^{\mu\nu}k_\mu k_\nu)^2
\end{equation} 
where
\begin{equation}
  \label{4.17}
\stackrel{1}{g}{}\!^{\mu\nu}=g^{\mu\nu}+(1-{1\over c_1^2})
\stackrel{\circ}{u}{}\!^{\mu}\stackrel{\circ}{u}{}\!^{\nu}
\end{equation}
and
\begin{equation}
  \label{4.18}
\stackrel{2}{g}{}\!^{\mu\nu}=g^{\mu\nu}+(1-{1\over c_2^2})
\stackrel{\circ}{u}{}\!^{\mu}\stackrel{\circ}{u}{}\!^{\nu}\ .
\end{equation}
Thus $k_\mu$ is characteristic for the second order system at natural
points 
 if either
$k_\mu$ is null w.r. to $\stackrel{1}{g}{}\!^{\mu\nu}$ or $k_\mu$ is null
w.r. to $\stackrel{2}{g}{}\!^{\mu\nu}$.
For the associated vectors $\alpha^A$ with $\Delta_{AB}\alpha^B=0$ we get
that $\stackrel{1}{\alpha}{}\!\!^A$ is proportional to
$\stackrel{\circ}{F}{}\!^A{}_\mu k^\mu$ and $\stackrel{2}{\alpha}{}\!^A$
is determined by $\stackrel{1}{\alpha}{}\!^A
\stackrel{2}{\alpha}{}\!^B G_{AB}=0$. Thus we have a longitudinal mode
propagating at phase velocity $c_1$ and two transversal modes propagating
at phase velocity $c_2$.

We now come to the question of symmetric hyperbolicity of
(\ref{3.7},\ref{3.8}), possibly after modifying
$\stackrel{\circ}{U}_{ACBD}$ by changing the factor of 2 in front of
$G_{A[C}G_{D]B}$ in (\ref{4.14})
into a generic constant $\sigma$. Setting $\sigma=4c_2^2-2\delta$ we find
the following 

{\bf Theorem 2:}  Let $0<\delta<{3c_1^2\over2}$ and $0<\delta<2c_2^2$.
Then the system (\ref{3.7},\ref{3.8}) is symmetric hyperbolic at natural
points.

{\bf Proof:} We show that $k_\mu=\stackrel{\circ}{u}_{\mu}$ is
timelike. Keeping in mind  (\ref{3.20}), this is equivalent to the
positivity of $S(\alpha,\alpha)$ given by
\begin{equation}
  \label{4.19}
S(\alpha,\alpha)=[
(c_1^2-2c_2^2)G_{AC}G_{BD}
+2c_2^2G_{D(C} G_{A)B} 
\end{equation}
$$
+(4c_2^2-2\delta) G{}_{A[C}G_{D]B}
]\alpha^{AC}\alpha^{BD}>0
$$
Decomposing
\begin{equation}
  \label{4.20}
\alpha^{AB}=\omega^{AB}+ \kappa^{AB} + {\kappa\over
3}G{}^{AB}
\end{equation}
where  $\omega^{AB}=\omega^{[AB]}$, $\kappa^{AB}=\kappa^{(AB)}$ and
$\kappa^{AB}G_{AB}=0$, we see that 
\begin{equation}
  \label{4.21}
S(\alpha,\alpha)=(c_1^2-{2\delta\over
3})\kappa^2+\delta\kappa_{AB}\kappa^{AB} +
(2c_2^2-\delta)\delta\omega_{AB}\omega^{AB} \ ,
\end{equation}
from which our assertion follows.

It follows from the above theorem that the system (\ref{3.7},\ref{3.8}) has
a well--posed initial value problem. But the allowable Cauchy data are
quite restricted. Namely they have to be near ones for which the matter
is initially static. The general allowable Cauchy data can be inferred
from the following
 
{\bf Theorem 3:} Suppose the equation of state $\epsilon(H^{AB})$ is of
the form (\ref{4.13}) with $0<{4\over3}c_2^2<c_1^2$. Then the system
(\ref{3.7},\ref{3.8}) with $0<\delta<2c_2^2$ is symmetric hyperbolic at
the natural points on the deformation bundle
$(\stackrel{\circ}{x},\stackrel{\circ}{X},\stackrel{\circ}{F})$ i.e.
ones where
$\stackrel{\circ}{F}{}\!^A{}_\mu\stackrel{\circ}{F}{}\!^B{}_\nu
g^{\mu\nu}
(\stackrel{\circ}{x})=G^{AB}(\stackrel{\circ}X)$. 
Covectors $k_\mu$ at $\stackrel{\circ}{x}$ are timelike iff
\begin{equation}
  \label{4.22}
\stackrel{1}{g}{}\!^{\mu\nu}k_\mu k_\nu<0\ \ \ {\rm and
}\stackrel{\circ}{u}{}^\mu k_\mu<0
\end{equation}
where $\stackrel{1}{g}{}\!^{\mu\nu}$ is given by (\ref{4.17}). We remark
that the additional hypothesis, i.e. that ${4\over3}c_2^2<c_1^2 $, is not
necessary for hyperbolicity but convenient for a complete
characterization of timelike covectors. This condition is however physically
entirely reasonable since it is equivalent to that the 'bulk modulus" given
by
$k={3\lambda+2\mu\over 3}$ be positive.

{\bf Proof:} Recall the definition (\ref{3.4}) of
$W^{\mu\nu}{}_{AB}{}^{(\lambda)}$ in terms of
$M^{\mu\nu}{}_{AB}$. In the case at hand $\bar M^{\mu\nu}{}_{AB}$
at natural points is given by 
\begin{equation}
  \label{4.23}
{1\over m\stackrel{\circ}{n}}\bar M^{\mu\nu}{}_{AB}=-G
_{AB}\stackrel{\circ}{u}\!{}^\mu\stackrel{\circ}{u}{}\!^\nu+
[
(c_1^2-2c_2^2)G_{AC}G_{BD}
+2c_2^2G_{D(C} G_{A)B} 
\end{equation}
$$
+(4c_2^2-2\delta) G_{A[C}G_{D]B}
]\stackrel{\circ}{F}{}\!^{C\mu}\stackrel{\circ}{F}{}\!^{D\nu}
$$
We have to study ${1\over m\stackrel{\circ}{n}}\bar W(k)={1\over
m\stackrel{\circ}{n}}\bar W^{\mu\nu}_{AB}{}^{(\lambda)}k_\lambda
m^B{}_\nu m^A{}_\nu$. Decomposing
\begin{equation}
  \label{4.24}
m^A{}_\mu=\alpha^A\stackrel{\circ}{u}_\mu+\stackrel{\circ}{F}{}\!^A{}_\mu\alpha^B{}_A
\end{equation}
and
\begin{equation}
  \label{4.25}
k_\mu=\omega \stackrel{\circ}{u}_\mu +\stackrel{\circ}{F}{}\!^A{}_\mu k_A
\end{equation}
there results
\begin{equation}
  \label{4.26}
{1\over m\stackrel{\circ}{n}}\bar W(k,k)=\omega\alpha_A\alpha^A +
(c_1^2-\delta)\omega\kappa^2+2(\delta-c_1^2)\kappa\alpha^Ak_A
+2c_2^2\omega\alpha_{AB}\alpha^{[AB]}
\end{equation}
$$
+\delta\omega\alpha_{AB}\alpha^{BA}+4c_2^2\alpha^Ak^B\alpha_{[AB]}-2\delta\alpha^Ak^B\alpha_{AB}\
.
$$
Here  indices are raised and lowered with $G
_{AB}$. In Appendix A it is shown that $\bar W$ is positive definite
iff
\begin{equation}
 \label{4.27}
 {\omega^2\over c_1^2}-k^Ak_A>0\ ,
\end{equation}
from which the Theorem follows.

We now take up the discussion of characteristic covectors of Sect.3. It
is easy to see from (\ref{3.19}) that the characteristic covectors, as
defined there for the 2nd--order system, are the same for
$M^{\mu\nu}{}_{AB}$ as for $\bar M^{\mu\nu}{}_{AB}$ . It remains to find
the characteristic covectors for the 1st--order system
(\ref{3.7},\ref{3.8}). Due to the block diagonal form of
(\ref{3.7},\ref{3.8}) the relevant determinant is given by
$-(u^\mu k_\mu)^3 D$, where $D$ is the (12 $\times$ 12) -- determinant
\begin{equation}
  \label{4.28}
D=det({1\over m\stackrel{\circ}n}\bar
W^{\mu\nu}{}_{AB}{}^{(\lambda)}k_\lambda)\ .
\end{equation}
The determinant,   in a frame adapted to
$\stackrel{\circ}u{}\!^\mu$, is ($a=c_1^2,\ b=c_2^2$):
$$
D=det\left(
{\begin{array}{c}
\rho \,\omega\,, \,0\,, \,0\,, \,a\,\mathit{k_1}\,, \,(\delta  - b)\,
\mathit{k_2}\,, \,(\delta  - b)\,\mathit{k_3}\,, \,b\,\mathit{k_2}\,
, \,(a - \delta )\,\mathit{k_1}\,, \,0\,, \,b\,\mathit{k_3}\,, \,0
\,, \,(a - \delta )\,\mathit{k_1} \\
0\,, \,\rho \,\omega\,, \,0\,, \,(a - \delta )\,\mathit{k_2}\,, \,b\,
\mathit{k_1}\,, \,0\,, \,(\delta  - b)\,\mathit{k_1}\,, \,a\,
\mathit{k_2}\,, \,(\delta  - b)\,\mathit{k_3}\,, \,0\,, \,b\,
\mathit{k_3}\,, \,(a - \delta )\,\mathit{k_2} \\
0\,, \,0\,, \,\rho \,\omega\,, \,(a - \delta )\,\mathit{k_3}\,, \,0\,, 
\,b\,\mathit{k_1}\,, \,0\,, \,(a - \delta )\,\mathit{k_3}\,, \,b\,
\mathit{k_2}\,, \,(\delta  - b)\,\mathit{k_1}\,, \,(\delta  - b)\,
\mathit{k_2}\,, \,a\,\mathit{k_3} \\
a\,\mathit{k_1}\,, \,(a - \delta )\,\mathit{k_2}\,, \,(a - \delta )
\,\mathit{k_3}\,, \,a\,\omega\,, \,0\,, \,0\,, \,0\,, \,a\,\omega - \delta 
\,\omega\,, \,0\,, \,0\,, \,0\,, \,a\,\omega - \delta \,\omega \\
(\delta  - b)\,\mathit{k_2}\,, \,b\,\mathit{k_1}\,, \,0\,, \,0\,, 
\,b\,\omega\,, \,0\,, \, - b\,\omega + \delta \,\omega\,, \,0\,, \,0\,, \,0\,, 
\,0\,, \,0 \\
(\delta  - b)\,\mathit{k_3}\,, \,0\,, \,b\,\mathit{k_1}\,, \,0\,, 
\,0\,, \,b\,\omega\,, \,0\,, \,0\,, \,0\,, \, - b\,\omega + \delta \,\omega\,, 
\,0\,, \,0 \\
b\,\mathit{k_2}\,, \,(\delta  - b)\,\mathit{k_1}\,, \,0\,, \,0\,, 
\, - b\,\omega + \delta \,\omega\,, \,0\,, \,b\,\omega\,, \,0\,, \,0\,, \,0\,, 
\,0\,, \,0 \\
(a - \delta )\,\mathit{k_1}\,, \,a\,\mathit{k_2}\,, \,(a - \delta )
\,\mathit{k_3}\,, \,a\,\omega - \delta \,\omega\,, \,0\,, \,0\,, \,0\,, \,a
\,\omega\,, \,0\,, \,0\,, \,0\,, \,a\,\omega - \delta \,\omega \\
0\,, \,(\delta  - b)\,\mathit{k_3}\,, \,b\,\mathit{k_2}\,, \,0\,, 
\,0\,, \,0\,, \,0\,, \,0\,, \,b\,\omega\,, \,0\,, \, - b\,\omega + \delta 
\,\omega\,, \,0 \\
b\,\mathit{k_3}\,, \,0\,, \,(\delta  - b)\,\mathit{k_1}\,, \,0\,, 
\,0\,, \, - b\,\omega + \delta \,\omega\,, \,0\,, \,0\,, \,0\,, \,b\,\omega\,, 
\,0\,, \,0 \\
0\,, \,b\,\mathit{k_3}\,, \,(\delta  - b)\,\mathit{k_2}\,, \,0\,, 
\,0\,, \,0\,, \,0\,, \,0\,, \, - b\,\omega + \delta \,\omega\,, \,0\,, \,b
\,\omega\,, \,0 \\
(a - \delta )\,\mathit{k_1}\,, \,(a - \delta )\,\mathit{k_2}\,, \,a
\,\mathit{k_3}\,, \,a\,\omega - \delta \,\omega\,, \,0\,, \,0\,, \,0\,, \,a
\,\omega - \delta \,\omega\,, \,0\,, \,0\,, \,0\,, \,a\,\omega
\end{array}}
\right)
$$
which gives (using Maple)
$$
D=  - \omega^{6}\,\delta ^{5}\,(2\,b - \delta )^{3}\,( -
2\,\delta 
 + 3\,a)\,(  - \omega^{2}\,\rho +b\,\mathit{ k_1}^{2}+b\,\mathit{k_2}^{2}
+ b\,\mathit{k_3}^{2} 
 )^{2} \\
(  - \omega^{2}\,\rho +a\,\mathit{k_1}^{2}+ a\,
\mathit{k_2}^{2} + a\,\mathit{k_3}^{2}        ) 
$$
or
\begin{equation}
  \label{4.29}
D=-\delta^5(u^\mu
k_\mu)^6(2c_2^2-\delta)^3(3c_1^2-2\delta)(\stackrel{1}{g}{}\!^{\mu\nu}k_\mu
k_\nu)(\stackrel{2}{g}{}\!^{\mu\nu}k_\mu k_\nu)^2\ .
\end{equation}
Note that $\Delta$ appears as a factor in $D$, as it has to be.

Finally we have to find Cauchy data. Such are given by a smooth
hypersurface $\Sigma\in M$ and on it data $X(x), F(x)$ satisfying conditions
(i,ii) of Theorem 1 so that the conormal $n_\mu$ of $\Sigma$ is everywhere
timelike in the sense of the symmetric hyperbolic system (\ref{S}). 
The easiest way to achieve this is as follows: Pick an arbitrary 
$\Sigma\subset M$  and
$\stackrel{\circ}{y}$ a point on $\Sigma$. Choose
$(\stackrel{\circ}{X},\stackrel{\circ}{F})$ so that
$G^{AB}(\stackrel{\circ}{X})=
\stackrel{\circ}{F}{}\!^A{}_\mu\stackrel{\circ}{F}{}\!^B{}_\nu
g^{\mu\nu}(\stackrel{\circ}{y})$ and, in addition, so that the conormal
$\stackrel{\circ}{n}{}_\mu$ of
$\Sigma$ at $\stackrel{\circ}{y}$ is timelike w.r. to 
$  \label{4.30}
\stackrel{1}{g}{}\!^{\mu\nu}=g^{\mu\nu}+(1-{1\over
c_1^2})\stackrel{\circ}{u}{}\!^{\mu}\stackrel{\circ}{u}{}\!^{\mu}$,
for $\stackrel{\circ}{u}{}\!^{\mu}$ given in terms of
$\stackrel{\circ}{F}{}\!^A{}_\mu$ by $\stackrel{\circ}{u}{}\!^{\mu}
={1\over
3!\stackrel{\circ}{n}}\epsilon^{\mu\nu\lambda\rho}\stackrel{\circ}{F}{}\!^A{}_\nu
\stackrel{\circ}{F}{}\!^B{}_\lambda\stackrel{\circ}{F}{}\!^C{}_\rho
\stackrel{\circ}\Omega_{ABC}$. 

Note that this condition is satisfied automatically when $c_1^2\leq1$ and 
$\Sigma$
is spacelike w.r. to $g_{\mu\nu}$ and can not be satisfied when 
$c_1^2>1$ and $\Sigma$ is 
timelike w.r. to $g_{\mu\nu}$. In the
other cases it can be satisfied by a suitable choice of
$\stackrel{\circ}{F}{}\!^A{}_\nu$. 

We now choose a function $\bar f:\Sigma \to B$  so that $\bar
f^A|_{\stackrel{\circ}{y}}=\stackrel{\circ}{X^A}$ and $\partial_i\bar
f^A|_{\stackrel{\circ}{y}}=\stackrel{\circ}F{}\!^A{}_i$. Choose $\bar
F^A{}_\mu$ on
$\Sigma$ so that $\bar F^A{}_\mu=\stackrel{\circ}F{}\!^A{}_\mu$ at
$\stackrel{\circ}y$ and $\bar F^A{}_i=\partial_i\bar f^A$ everywhere.
Then, in a sufficiently small neighbourhood
$O$ of $\stackrel{\circ}y$, the field equations (\ref{3.1}) with initial data
$f^A=\bar f^A$ and $\partial_\mu f^A=\bar F^A{}_\mu$ on $\Sigma$ have a
unique local solution.

\section{Discussion}
\setcounter{equation}{0}
We want to discuss certain physical situations in which our results show
existence and uniqueness of solutions of the elasticity equations. Let us
stress that in the present paper we are concerned only with local
questions. In our formulation of elasticity theory 
we have as yet no way to solve the
boundary initial value problem corresponding to the motion of
a finite elastic body, i.e. where the normal component of the stress
tensor at the (free!) boundary of the body is required to be zero.
Hence we have to consider just parts of
the body or infinitely extended bodies. 

First we consider elasticity in Special Relativity. Hence our spacetime
metric is Minkowski space, i.e. $g_{\mu\nu}=\eta_{\mu\nu}$. Furthermore
we assume that the coordinates are inertial, then $t=const$ is the
natural Cauchy surface. We assume the body metric to be flat, i.e.
$G^{AB}=\delta^{AB}$. The map $\stackrel{\circ}f:
(t,x^1,x^2,x^3)\mapsto(X^1=x^1,X^2=x^2,X^3=x^3)$ is a solution for the
elasticity equation if the Lagrangian is given by (\ref{4.13}). The
interpretation of this solution is that of  a relaxed body at rest at all
times. The obvious physical question is to deform the body slightly at
$t=0$ and ask for solutions. Any map $f$ sufficiently near to
$\stackrel{\circ}f$ at {t=0} defines data $f^A,F^A{}_\mu$ at $t=0$
satisfying the positivity requirement via Theorem 3 in Section 4. 
We thus obtain
existence locally in time for pieces of the body 
or for an infinitely extended body. 

Consider next a general spacetime to be given and in it a spacelike
hypersurface $\Sigma$. If the geometry on and near $\Sigma$ is "close"
 to that of Minkowski space and the data close to the ones in the
previous paragraph, we obtain again a solution local in time. 

In an arbitrary spacetime we can, following the procedure at the end of
Sect 4, construct solutions locally both in space and time. 

Consider as a further example a spacetime of the form
\begin{equation}
  \label{5.1}
ds^2=-dt^2+ g_{ik}(t,x^j)dx^i dx^k
\end{equation}  
Let $\psi$ be a diffeomorphism from $t=0$ onto the body and define the
body metric $G{}\!^{AB}$ such that $\psi$ becomes an
isometry. Then the system becomes symmetric hyperbolic for the data
$f^A=\psi^A, \partial_\mu f^A=0$ at $t=0$, or data near-by. We obtain
solutions globally in space, locally in time.

{\bf{Acknowledgments}}: We thank John Ball for very helpful suggestions and
J\"urgen Ehlers for his detailed comments on the manuscript.

\begin{appendix}
\section{Appendix A}
\setcounter{equation}{0}
We start with the expression (\ref{4.23}) and decompose $\alpha^{AB}$ in
the form
\begin{equation}
  \label{A1}
\alpha^{AB}=\omega^{AB}+ \kappa^{AB} + {\kappa\over
3}G^{AB}
\end{equation}
where  $\omega^{AB}=\omega^{[AB]}$, $\kappa^{AB}=\kappa^{(AB)}$ and
$\kappa^{AB}G_{AB}=0$. 
There results 
\begin{eqnarray}
\label{A2}
{1\over m\stackrel{\circ}n}\bar W(k)(m,m)&=&
\omega\alpha_A\alpha^A+2(\delta-c_1^2)\alpha^Ak_A\kappa\\ 
  & & +2\omega
c_2^2\omega_{AB}\omega^{AB}+4c^2_2\omega_{AB}\alpha^Ak^B\nonumber\\
  & & -2\delta(\omega_{AB}+\kappa_{AB}+{\kappa\over 3}h_{AB})\alpha^Ak^B\nonumber\\
  & &+
(c_1^2-\delta)\omega\kappa^2+\delta\kappa(-\omega_{AB}\omega^{AB}+\kappa_{AB}\kappa^{AB}+
{\kappa^2\over 3})\nonumber 
\end{eqnarray}
We now proceed as follows: we first eliminate all linear  $k_A$ -- terms by
substituting $\alpha_A $ with $\beta_A$ given by 
\begin{equation}
  \label{A3}
\beta_A=\alpha_A+{1\over\omega}({2\delta\over3}-c_1^2)\kappa k_A+
{1\over\omega}(2c_2^2-\delta)\omega_{AB}k^b
-{\delta\over\omega}\kappa_{AB}k^B
\end{equation}
Next we eliminate terms linear in $\kappa_{AB}$ by setting
\begin{eqnarray}
\label{A4}
\bar\kappa_{AB}&=&A\kappa_{AB}-{\delta\kappa\over
A\omega}(c_1^2-{2\delta\over3})k_Ak_B\\ 
&&+{\delta\kappa k_Ck^C\over A\omega}(c_1^2-{2\delta\over 3}){1\over
3}h_{AB}\nonumber\\ &&+{\delta\over
A\omega}(2c_2^2-\delta)k_{(A}\omega_{B)C}k^C\nonumber
\end{eqnarray}
for some constant $A\neq0$. Thirdly, note the identity
\begin{equation}
  \label{A5}
2\bar\kappa_{ABC}\bar\kappa^{ABC}=\kappa_{AB}\kappa^{AB}k_Ck^C-{3\over
2}\kappa_{AB}k^B\kappa^A{}_Ck^C
\end{equation}
where $\bar\kappa_{ABC}$ is the trace--free part of $\kappa_{A[B}{}k_{C]}$, i.e.
\begin{equation}
  \label{A6}
\bar\kappa_{ABC}:=\kappa_{A[B}{}k_{C]}+{1\over 2}h_{A[B}\kappa_{C]D}k^D
\end{equation}
We use (\ref{A6}) to eliminate $\kappa_{AB}k^B\kappa^A{}_Ck^C$ in favour of the
other two quantities in (\ref{A5}). We do the analogous thing for
$\omega_{AB}$, using
\begin{equation}
  \label{A7}
3\omega_{[AB}{}k_{C]}\omega^{[AB}{}k^{C]}=\omega_{AB}\omega^{AB}k_Ck{}^C
-\omega^A{}_Bk^B\omega_{AC}k^C
\end{equation}
The last two
operations leave us with just having to worry about the signs of the terms
proportional to $\kappa^2$, $\kappa_{AB}\kappa^{AB}$ and $\omega_{AB}\omega^{AB}$.
We have to choose $A$ such that all the signs are positive and, for the "only if"
--direction, have to make an optimal choice in terms of the allowed range for
$k_Ak^A$. Our choice is $A={\delta\omega\over
c_1^2}(c_1^2-{2\delta\over3})$. We finally obtain
\begin{eqnarray}
\label{A8}
{1\over m\stackrel{\circ}n}\bar
W(k)(m,m)&=&\omega\beta_A\beta^A+
\delta\omega(1-{2\delta\over3c_1^2})\bar\kappa_{AB}\bar\kappa^{AB}\\
 &&{2\delta^2\omega\over3}({1\over
c_1^3}-{k_Ck^C\over\omega^2})\kappa_{AB}\kappa^{AB}\nonumber\\ 
&&+{2\delta\omega c_1^2\over3}(c_1^2-{2\delta\over3})({1\over
c_1^2}-{k_Ck^C\over\omega^2})({3\over2\delta}
+{k_Dk^D\over\omega^2})\kappa^2\nonumber\\
&&+\omega(2c_2^2-\delta)D({k^Ak_A\over\omega^2})\omega_{AB}\omega^{AB}\nonumber\\
&&+{3(2c_2^2-\delta)\over2\omega}\left[1+{\delta
c_1^2\over2\omega^2(c_1^2-{2\delta\over3})}
\right]\omega_{[AB}k_{C]}\omega^{[AB}k^{C]}\nonumber\\
&&+{4\delta^2\over3\omega}\bar\kappa_{ABC}\bar\kappa^{ABC}
\nonumber
\end{eqnarray}
where
\begin{equation}
  \label{A9}
D(\eta)=1-{2c_2^2-\delta\over2\omega^2}\eta-{(2c_2^2-\delta)\delta
c_1^2\over4\omega^2(c_1^2-{2\delta\over3})}\eta^2
\end{equation} 
Observe that $D(0)>0$ and 
\begin{equation}
  \label{A10}
D({1\over c_2^2})={\delta^2\over
4c_2^2}
{c_1^2-{4\over3}c_2^2\over c_1^1-{2\delta\over3}}>0
\end{equation}
Consequently $D(\eta)$ is positive for all $\eta\in[0,{1\over c_1^2}]$, since 
${1\over c_1^2}<{1\over c_2^2}$ and $D(\eta)$ is monotonically decreasing for
$\eta>0$. Thus for ${1\over c_1^2}-{k_Ak^A\over \omega^2}>0$, $\bar W(k)(m,m)$
is positive, except possibly when $\omega_{AB}, \bar\kappa_{AB},\kappa$ and
$\beta_A$ are all zero. But then $\kappa_{AB}=0=\alpha_A$.

Suppose, conversely, that ${1\over c_1^2}-{k_Ak^A\over \omega^2}\leq0$. Then
choose $\omega_{AB}=0$, $\kappa$ non--zero, $\kappa_{AB}$ such that
$\bar\kappa_{AB}=0$ and $\alpha_A$ such that $\beta_A=0$. Clearly
$\kappa_{AB}$ is non--zero of the form "trace--free part of $k_Ak_B$'',whence
the last line in (\ref{A8}) is zero,as are the first and fourth line. But the
second and third line are non--positive. Thus, (\ref{A2}) is positive definite
iff ${\omega^2\over c_1^2}-k_Ak^A>0$.

\section{{\bf Appendix B}}
\setcounter{equation}{0}
Here we outline how nonrelativistic elasticity is recovered by taking the
limit $c\to\infty$ in our equations when $(M,g)$ is the Minkowski space
\footnote{A related discussion can be found in \cite{So}.}.
We have 
\begin{equation}
  \label{B1}
g_{\mu\nu}dx^\mu dx^\nu=-c^2 dt^2 +\delta_{ik}dx^idx^k
\end{equation}
\begin{equation}
  \label{B2}
g^{\mu\nu}\partial_\mu\partial_\nu=-{1\over c^2} (\partial_t)^2
+\delta^{ik}\partial_i\partial_k
\end{equation}
and
\begin{equation}
  \label{B3}
u^\mu\partial_\mu={1\over c}{1\over\sqrt{1-{v^2\over
c^2}}}(\partial_t+v^i\partial_i) 
\end{equation}
\begin{equation}
  \label{B4}
u_\mu dx^\mu={1\over\sqrt{1-{v^2\over
c^2}}}(-cdt+{v_i\over c}dx^i) 
\end{equation}
where $v_i=\delta_{ik}v^k$ and $v^2=v_iv^i$. The vector $v^i$ is
determined from $F^A{}_\mu dx^\mu=F^Adt+f^A{}_idx^i$ by  
\begin{equation}
  \label{B5}
F^A+F^A{}_iv^i=0
\end{equation}
The tensor $h^\mu{}_\nu=\delta^\mu_\nu+u^\mu u_\nu$ goes in the linit
$c\to\infty$ to $\delta^\mu_\nu-v^\mu\tau_\nu$ where
\begin{equation}
  \label{B6}
v^\mu\partial_\mu=\partial_t+v^i\partial_i
\end{equation}
and
\begin{equation}
  \label{Bt}
\tau_\mu dx^\mu=dt
\end{equation}
The quantity $F_A{}^\mu\partial_\mu$ tends to $F_A{}^i\partial_i$ and
\begin{equation}
  \label{B7}
F_A{}^\mu F^A{}_\nu=\delta^\mu_\nu-v^\mu\tau_\nu\ ,\ F^A{}_\mu
F_A{}^\mu=\delta^A_B
\end{equation}
The quantity $H^{AB}=F^A{}_\mu F^B{}_\nu g^{\mu\nu}$ tends to
\begin{equation}
  \label{B8}
K^{AB}=F^A{}_i F^B{}_j\delta^{ij}
\end{equation}
Finally, the equation of state $\epsilon(X^A, H^{BC})$  can 
not be expected to
have a finite limit as $c\to\infty$, but $e(X^A, H^{BC})$ has 
a limit, where
\begin{equation}
  \label{B9}
\epsilon= m c^2+e(H^{AB})=m c^2+e(K^{AB})+ O({1\over c^2})
\end{equation}
We omit the term $m c^2$ ( which diverges as $c\to\infty$) from  the
Lagrangian density (\ref{B9}) because it contributes $nm c^2$ to the
Lagrangian density which has vanishing Lagrangian derivative and thus
leaves the field equations unchanged. We insert (\ref{4.8},\ref{4.9}) into
(\ref{3.18}) and take the limit
$c\to \infty$ to finally obtain
\begin{equation}
  \label{B10}
M^{\mu\nu}{}_{AB}=-nm K_{AB}\  v^\mu v^\nu+n\left[
\tau_{AB}K_{CD} + \tau_{AC}K_{BD}+\tau_{BD} K_{AC} +2
{\partial\tau_{AC}\over\partial K^{BD}}
\right] F^{C\mu}F^{D\nu}\ ,
\end{equation}
where $F^{A\mu}=F^A{}_\mu h^{\mu\nu}$ with
\begin{equation}
  \label{B11}
h^{\mu\nu}\partial_\mu\partial_\nu=\delta^{ik}\partial_i\partial_k\ .
\end{equation}
Furthermore $\tau_{AB}=2{\partial e\over \partial K^{AB}}$ and $n$ is
defined by
\begin{equation}
  \label{B12}
6n^2=K^{AA'}K^{BB'}K^{CC'}\Omega_{ABC}\Omega_{A'B'C'}\ , n>0
\end{equation}
Thus the structure of the nonrelativistic equations is very similar to
that of the relativistic ones. In particular, we can obtain  existence
(locally in time) by essentially the method described in the  body of the
paper.

In nonrelativistic elasticity it is common to use as the basic field
variable $\phi^i(t,X)$ defined by 
\begin{equation}
  \label{B13} 
f^A(t,\phi(t,X))=X^A
\end{equation}
We state without proof that the equations derived above, when rewritten in
terms of $\phi$, are equivalent to the ones usually considered in the
frame-indifferent, hyperelastic case (see Gurtin \cite{Gu}).
\end{appendix}

\end{document}